\documentclass[12pt]{article}
\usepackage{graphicx,subfigure}
\usepackage{epsfig}
\usepackage{mathrsfs}
\usepackage{amsmath}
\usepackage{amsfonts}
\usepackage{amssymb}
\usepackage[usenames]{color}
\usepackage[a4paper,left=2.5cm,right=2.5cm,top=2.5cm,bottom=2.5cm]{geometry}
\usepackage{hyperref}

\def\be{\begin{equation}}
\def\ee{\end{equation}}
\def\bea{\begin{eqnarray}}
\def\eea{\end{eqnarray}}

\begin{document}

\makeatletter
\renewcommand{\theequation}{\thesection.\arabic{equation}}
\@addtoreset{equation}{section}
\makeatother

\baselineskip 18pt

\def\todo#1{\textbf{[}\emph{\textbf{#1}}\textbf{]}}

\begin{titlepage}

\vfill

\begin{flushright}
\end{flushright}

\vfill

\begin{center}
   \baselineskip=16pt
   {\Large\bf The moduli space of striped black branes}
  \vskip 1.5cm
      Benjamin Withers\\
   \vskip .6cm
     \textit{Centre for Particle Theory and Department of Mathematical Sciences\\ Durham University, South Road, Durham, DH1 3LE, U.K.}


\end{center}

\vfill

\begin{center}
\textbf{Abstract}
\end{center}

\begin{quote}
At finite charge density certain holographic models exhibit the spontaneous breaking of translational invariance resulting in an inhomogeneous phase. We follow up on recent numerical work, reporting results for a larger class of cohomogeneity two black branes in AdS, dual to a holographic striped phase. We construct the continuous moduli space of inhomogeneous black branes as a function of the temperature. Minimising the free energy we determine the dominant striped solutions, revealing a growth in the stripe size as the system is cooled. We discuss the thermodynamic properties of this line of solutions.
\end{quote}

\vfill

\end{titlepage}

\section{Introduction}
There is great interest in applying the holographic framework to systems which may be of relevance to real-world condensed matter phenomena, with one focus placed on equilibrium phases at finite charge density and temperature.
An early example is provided by the holographic description of a superfluid, which can be built with the minimal addition of a charged scalar field to the bulk gravitational theory. The scalar may be added in a phenomenological fashion \cite{Gubser:2008px,Hartnoll:2008vx,Hartnoll:2008kx} or by seeking appropriate consistent truncations of supergravity reductions \cite{Gauntlett:2009dn,Gubser:2009qm}. The superfluid phase transition in the CFT is described in the bulk by a branch of thermodynamically preferred black branes with scalar condensate connecting with the normal phase.

There is growing evidence that symmetry breaking phases may be generic in holography at finite charge density. In particular, linear analyses indicate the spontaneous breaking of the Euclidean group of symmetries \cite{Nakamura:2009tf,Ooguri:2010kt,Donos:2011bh,Donos:2011ff} in a variety of gravitational settings \cite{Vegh:2013sk,Donos:2013gda,Cremonini:2012ir,Iizuka:2013ag} as well as probe brane constructions \cite{Domokos:2007kt,Ooguri:2010xs,Bayona:2011ab,Bergman:2011rf}. To establish whether a phase transition exists, construction of the broken solutions beyond linear order is necessary, which typically requires the solution of PDEs.  In cases where the broken phase retains homogeneity, fully backreacted solutions have been constructed by solving ODEs \cite{Iizuka:2012iv,Donos:2012gg,Donos:2012wi} exhibiting the emergence of a thermodynamically preferred spatial scale, $\lambda_k \equiv 2\pi/k$ associated with helical order. Such solutions provide an important set of examples for comparison with inhomogeneous solutions. 


Recently the first examples showing continuous phase transitions to an inhomogeneous phase in this context were presented in \cite{DonosSolo,Withers:2013loa} (see also \cite{UBCv1}), in the Einstein-Maxwell-pseudoscalar bulk models studied perturbatively in~\cite{Donos:2011bh}.  There, a branch of solutions at fixed periodicity $\lambda_k\mu$ was considered, connecting with a single striped zero-mode of the RN background with wavenumber $k$. In this paper we relax the condition of fixed $k/\mu$, finding a continuous moduli space of striped solutions at each temperature. In particular, we consider the natural generalisation of the solutions constructed in \cite{DonosSolo,Withers:2013loa}, by seeking the nonlinear one-parameter families which emerge from a given striped zero-mode, $k$, for all $k$. In this way we construct a two-parameter family of solutions labelled by the dimensionless temperature, $T/\mu$, and periodicity scale, $k/\mu$. 

For an infinite translationally invariant system at temperature $T/\mu$, the physically relevant solutions in this space are those which minimise the free energy. These can be labelled by their periodicity, giving rise to the curve  $\tfrac{k}{\mu}(\tfrac{T}{\mu})$. 
Considering this line of solutions we find a second order phase transition from the homogeneous phase. In the helical case \cite{Donos:2012gg,Donos:2012wi} it was found that the dominant scale at any given temperature was characterised by the vanishing of a particular piece of boundary data. We will uncover an analogous result for  inhomogeneous solutions constructed here, in one lower dimension.

\emph{Note added:} The paper \cite{UBCv2}, an update of \cite{UBCv1}, has recently appeared indicating the dominant scale in the canonical ensemble for a similar model. Continuous phase transitions, as found in \cite{DonosSolo,Withers:2013loa}, were also found the updated version.
 
\section{The setup of \cite{Withers:2013loa}\label{setup}}
In this section we briefly recap the model and numerical setup of \cite{Withers:2013loa}.
We adopt a bulk model studied in \cite{Donos:2011bh} containing a neutral pseudo-scalar, $\phi$, a single $U(1)$ gauge field, $\mathcal{A}$ with field strength $F=d\mathcal{A}$, and crucially in this context, a parity-violating coupling, $\vartheta(\phi)F\wedge F$,
\be
{\scriptsize
S_{\text{b}} = \int d^4x \sqrt{-g}\left(R -\frac{1}{2}\left(\partial \phi\right)^2 - \frac{\tau(\phi)}{4}F^2 - V(\phi)\right)- \int \frac{\vartheta(\phi)}{2}F\wedge F,\label{sbulk}}
\ee
where we have set $16\pi G=1$.
In this paper we study a single case, taking $\tau = \text{sech}(\sqrt{3}\phi)$, $V = -6\,\text{cosh}(\phi/\sqrt{3})$ and $\vartheta = \frac{c_1}{6\sqrt{2}}\,\text{tanh}(\sqrt{3}\phi)$. When $c_1=6\sqrt{2}$ this phenomenological model becomes a consistent truncation of a reduction from 11D on SE$_7$ \cite{Gauntlett:2009bh,Gauntlett:2009zw,Donos:2012yu}. Here we adopt the slightly higher value of $c_1=9.9$ which raises the critical temperature for the striped instability.
We seek regular stationary solutions within the bulk metric ansatz 
\be
ds^2 = \frac{1}{z^2}\big(-T f(z)dt^2 + Z\frac{dz^2}{f(z)}+X (dx+\gamma dz)^2 + Y(dy+\beta dt)^2\big)\label{ba1}
\ee
and gauge field $\mathcal{A} = A dt + B dy$, where $T,Z,X,Y,\gamma,\beta, A, B$ and $\phi$ are functions both of the AdS radial coordinate, $z$, and a single spatial boundary direction $x$ with translational invariance in $y$. 
%
The function $f(z) \equiv (1-z)(1+z+z^2-\mu^2z^3/4)$ conveniently factors out the normal phase solution, an electrically charged Reissner-Nordstrom (RN) black brane branch at $Z=T=X=Y=1$ and $\gamma=\beta=B=\phi=0$ and 
$A= \mu(1-z)$. The coordinate $z$ runs from $z=0$ at the boundary to $z=1$ at the (non-degenerate) horizon. To render the system elliptic we adopt the Harmonic Einstein equation approach of  \cite{Headrick:2009pv,Adam:2011dn,Wiseman:2011by} and we employ a spectral method with $N\times (N+1)$ grid points in the bulk. Further details can be found in \cite{Withers:2013loa}.
For the data presented in this paper we have taken $N=24$ for $0.01\leq T/\mu\leq(T/\mu)_{c,\text{max}}$, $N=30$ for $0.005\leq T/\mu<0.01$, $N=40$ for $0.003 \leq T/\mu<0.005$, $N=50$ for $0.0015\leq T/\mu<0.003$ and finally $N=52$ for $T/\mu=0.001$.

The non-vanishing components of the expectation value of the CFT stress tensor and current are given by, 
\bea
\varepsilon\equiv \left<T_{tt}\right> &=& 2+\frac{\mu^2}{2}-3T^{(3)}(x)\\
\left<T_{ty}\right> &=& 3 \beta^{(3)}(x), \\
\mathcal{P}_x\equiv\left<T_{xx}\right> &=& 1+\frac{\mu^2}{4} + 3 X^{(3)}\\
 \mathcal{P}_y\equiv\left<T_{yy}\right> &=& 1+\frac{\mu^2}{4} - 3 X^{(3)} - 3T^{(3)}(x),
 \eea
 \bea
 \rho\equiv \left<J_t\right> & =&-A^{(1)}(x)\\
  \left<J_y\right> &=& -B^{(1)}(x).
\eea
where $F^{(n)}$ denotes the coefficient of $z^{n}$ in the small-$z$ boundary expansion of the field $F$. Note that the stress tensor is traceless and $X^{(3)}$ is constant in $x$, following from conservation of the stress tensor. Spatial averages of these quantities will be denoted with a bar.

\section{Striped zero-modes\label{zeromodes}}
We begin with a summary of the striped zero-modes on the RN normal phase solution, following \cite{Donos:2011bh}. These modes indicate the boundary of stability for the RN solution, providing a starting point for the construction of an emergent family of nonlinear striped black branes. Moreover, in this model at any temperature the boundary of stability will coincide with the boundary of existence for the striped solutions (labelled by their periodicity), as we shall show in the following section. 

\begin{figure}[h!]
\begin{center}
\includegraphics[width=0.5\columnwidth]{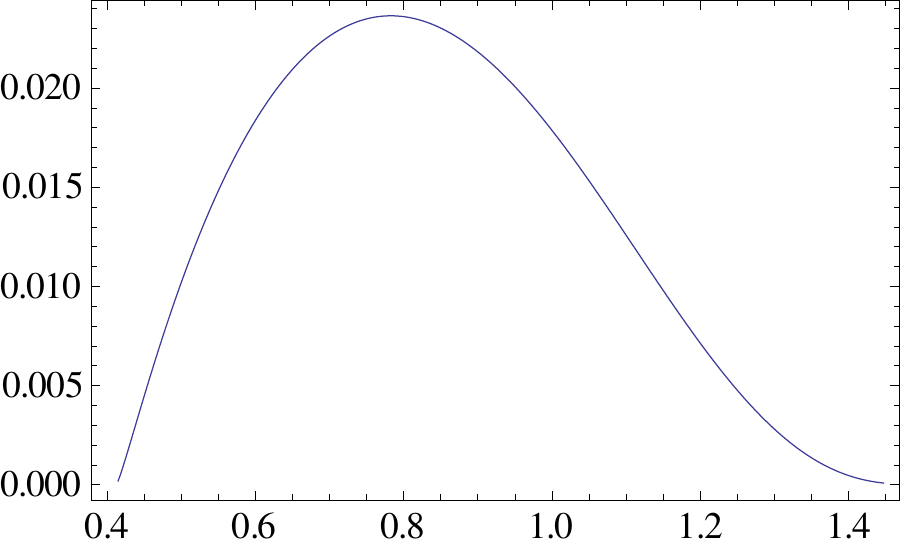}
\begin{picture}(0.1,0.1)(0,0)
\put(-250,80){\makebox(0,0){$T/\mu$}}
\put(-110,-10){\makebox(0,0){$k/\mu$}}
\end{picture}
\vskip 1em
\caption{Critical temperatures for the onset of the striped instability for each wavenumber $k/\mu$, constructed using a zero-mode analysis. For the model considered in this paper this curve gives the boundary of the region where striped solutions exist.\label{bellcurve}}
\end{center}
\end{figure}
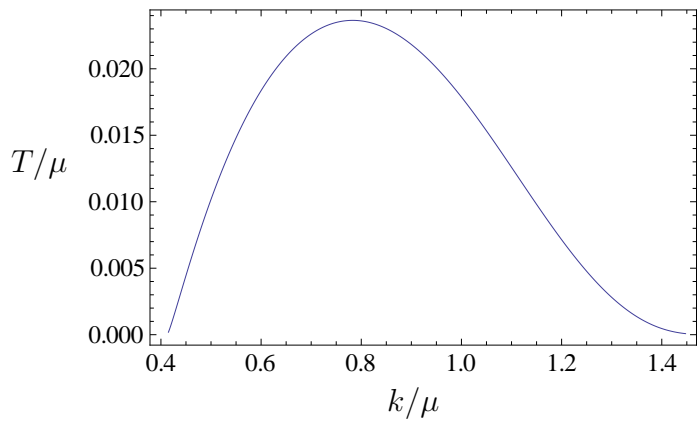

Consider a single Fourier mode with wavenumber $k$ for the following perturbations,
\be
\phi = \epsilon\, \phi_k(z) \cos{kx},\qquad B = \epsilon\, B_k(z)\sin{kx}, \qquad\beta = \epsilon\, \beta_k(z)\sin{kx},
\ee
with the others set at their RN values. The resulting equations at $\mathcal{O}(\epsilon)$ are three second order k-dependent linear ODEs. With the conditions of horizon regularity and normalisability in the UV, counting the number of pieces of boundary data reveals that there can be at most discrete striped zero-modes at a given $T/\mu$. In particular for $T/\mu<(T/\mu)_{c,\text{max}}\simeq 0.0236$ there are two zero-modes at fixed $T/\mu$, giving rise to the characteristic critical temperature `bell curve' presented in figure \ref{bellcurve}. 

As the system is cooled we first encounter a single zero-mode at $T/\mu = (T/\mu)_{c,\text{max}}$ at the scale $k/\mu \simeq 0.783$. Thus the natural strategy for constructing nonlinear solutions is to consider those solutions which are continuously connected to this mode. In particular, in the next section we consider those solutions which emerge from starting with a single zero-mode of wavenumber $k$ and lowering the temperature. This gives rise to a two parameter family of solutions labelled by $k/\mu$ and $T/\mu$.

As a side remark, as there are two zero-modes at temperatures $T/\mu < (T/\mu)_{c,\text{max}}$, a second possibility is that there are additional branches of nonlinear solutions which connect with a superposition of them. In the numerical method used we must fix the periodicity of the solution and so this scenario is largely excluded, except at a discrete set of temperatures for which the higher zero-mode wavenumber is an integer multiple of the lower. We have not constructed any solutions of this type, though we anticipate that if they do exist they will be thermodynamically subdominant.

\section{Nonlinear solutions\label{nonlinear}}
If we fix the periodicity of the solution and lower the temperature then \cite{Withers:2013loa} (see also \cite{DonosSolo} for similar models) showed the existence of a second order phase transition. More generally though, as discussed in section \ref{zeromodes}, there is a two parameter family of solutions connecting with the RN branch along the line of striped zero-modes shown in figure \ref{bellcurve}. Here we construct these solutions and investigate the thermodynamically preferred stripe as a function of the temperature, $\tfrac{k}{\mu}(\frac{T}{\mu})$.

\begin{figure}[h!]
\begin{center}
\includegraphics[width=0.45\columnwidth]{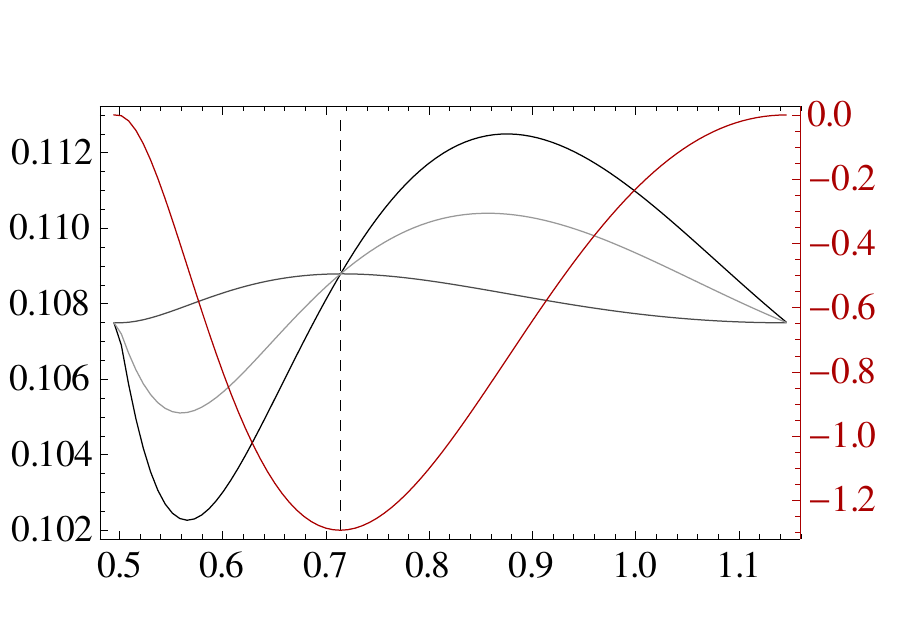}
\hspace{0.05\columnwidth}
\includegraphics[width=0.45\columnwidth]{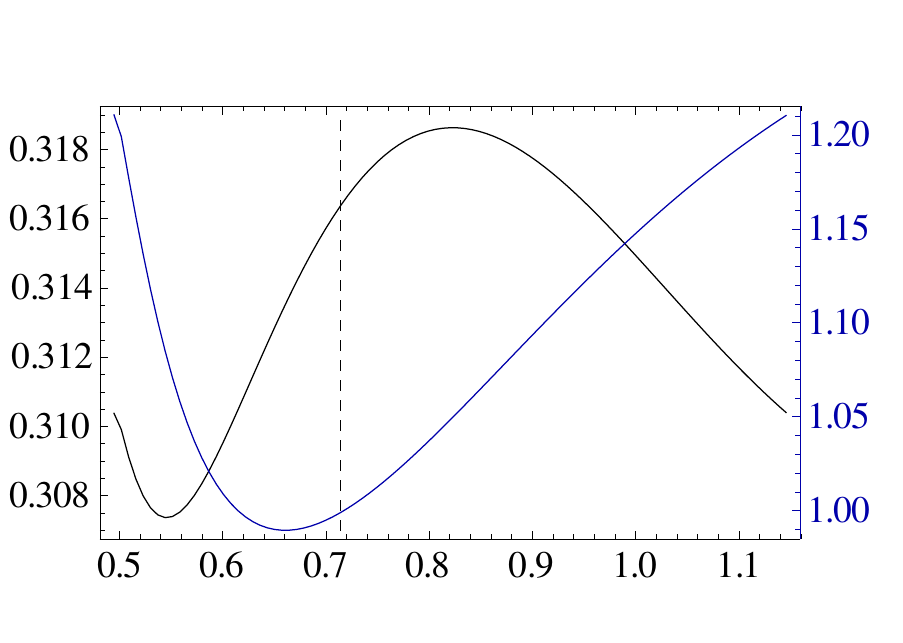}
\begin{picture}(0.1,0.1)(0,0)
\put(-110,0){\makebox(0,0){$k/\mu$}}
\put(-350,0){\makebox(0,0){$k/\mu$}}
\put(-350,115){\makebox(0,0){\scriptsize $\mathcal{P}_x/\mu^3$}}
\put(-350,76){\makebox(0,0){\scriptsize $\bar{\mathcal{P}}_y/\mu^3$}}
\put(-330,102){\makebox(0,0){\scriptsize $\bar{\varepsilon}/2\mu^3$}}
\put(-320,40){\makebox(0,0){\scriptsize $10^3\frac{\bar{w}-w_{\text{RN}}}{\mu^3}$}}
\put(-90,100){\makebox(0,0){\scriptsize $\bar{\rho}/\mu^2$}}
\put(-90,50){\makebox(0,0){\scriptsize $\bar{s}/\mu^2$}}
\end{picture}
\vskip 1em
\caption{The spatially averaged stress tensor, free energy, charge density and entropy density in the space of solutions labelled by $k/\mu$ a fixed $T/\mu = 0.01$. The dashed line indicates the $k/\mu$ which minimises $\bar{w}$ at this temperature.\label{fixedT}}
\end{center}
\end{figure}
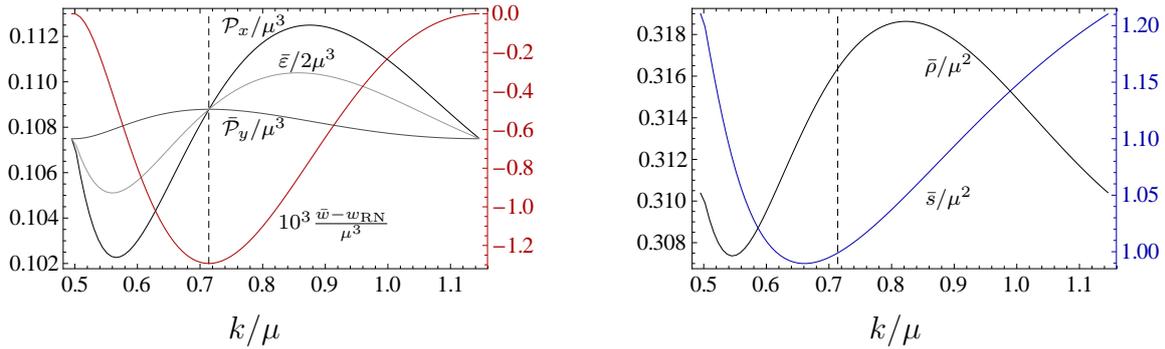

We begin by considering the space of such solutions which exist at a fixed temperature $T/\mu$, for which various averaged thermodynamic quantities are shown in figure \ref{fixedT}. First we note that the striped solutions exist within the interval defined by the locations of the two zero-modes at this temperature. Additionally we see that the free energy and the entropy of the striped phase is always lower than the RN value. Note that the thermodynamically preferred striped solution at this temperature appears to have the property $\mathcal{P}_x = \bar{\mathcal{P}}_y$, and we will discuss this shortly. The relationship between $\bar{\varepsilon}$ and $\mathcal{P}_x, \bar{\mathcal{P}}_y$ follows from conformality.

\begin{figure}[h!]
\begin{center}
\includegraphics[width=0.6\columnwidth]{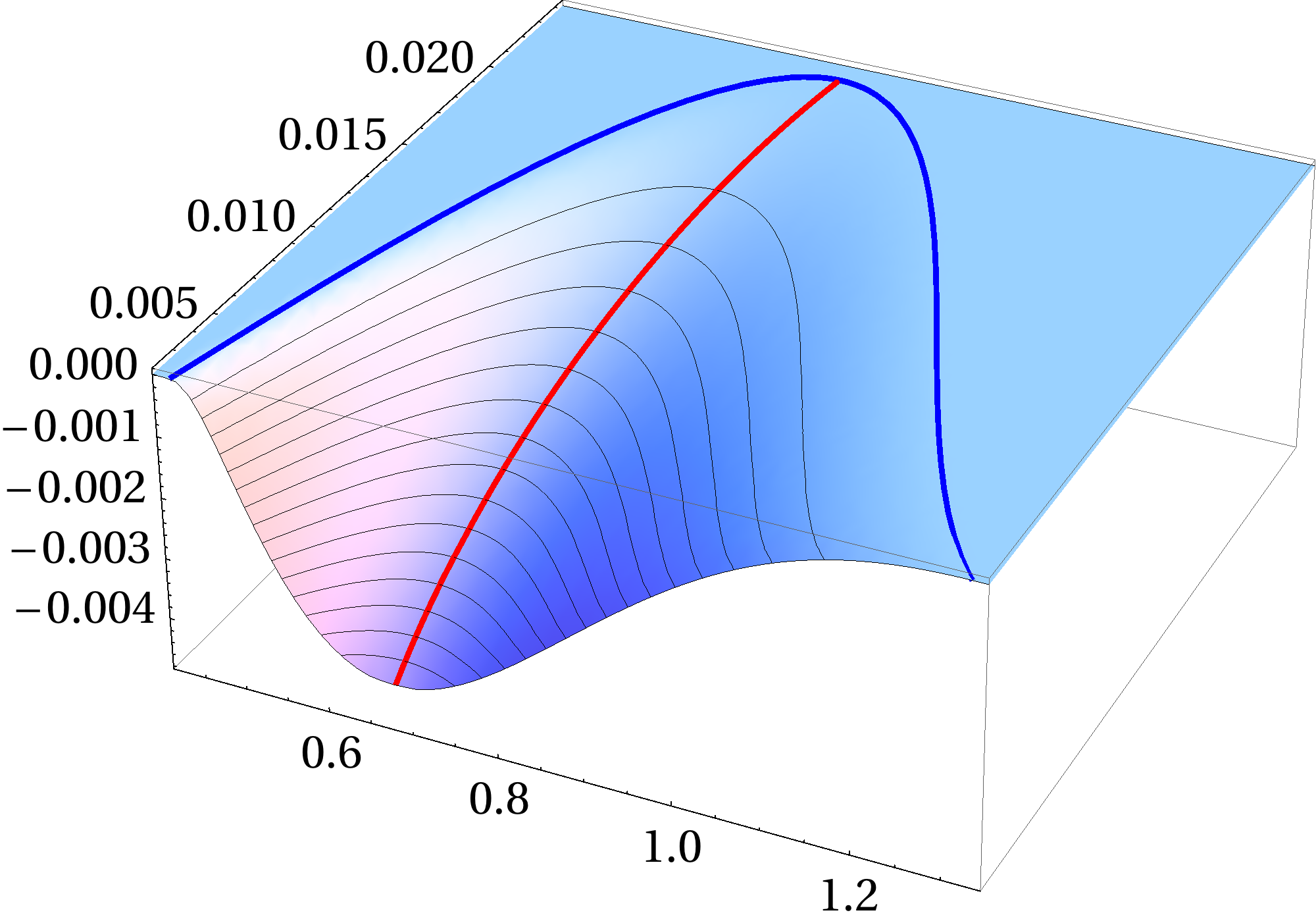}
\begin{picture}(0.1,0.1)(0,0)
\put(-172,8){\makebox(0,0){$k/\mu$}}
\put(-240,168){\makebox(0,0){$T/\mu$}}
\put(-293,95){\makebox(0,0){$\frac{\bar{w}-w_{\text{RN}}}{\mu^3}$}}
\end{picture}
\vskip 1em
\caption{The two parameter family of striped black branes, which exist at temperatures below the line of striped zero-modes indicated in blue. Plotted is the difference between the spatially-averaged free energy $\bar{w}(k/\mu,T/\mu)$ and that of the RN solution, with the striped solutions dominant. The red line indicates the line of solutions obtained by minimising $\bar{w}$ with respect to $k/\mu$ at constant $T/\mu$. The remaining contours show solutions of equal free energy difference. \label{free3d}}
\end{center}
\end{figure}
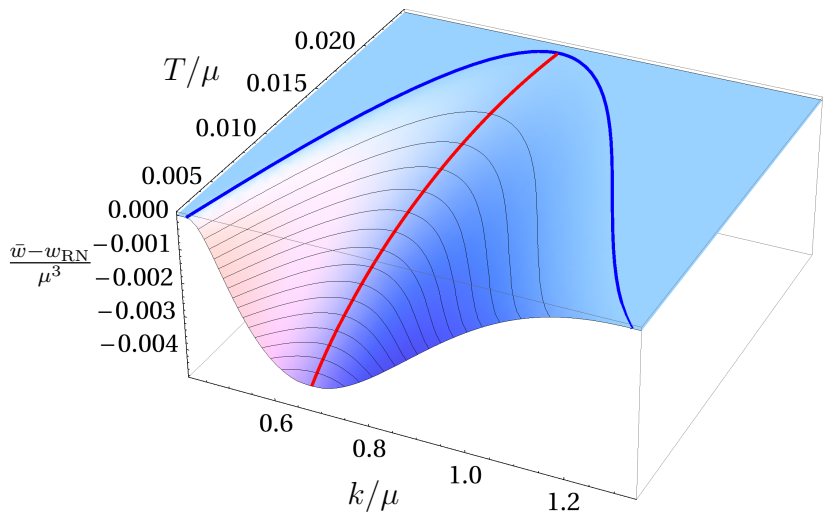

\begin{figure}[h!]
\begin{center}
\includegraphics[width=0.5\columnwidth]{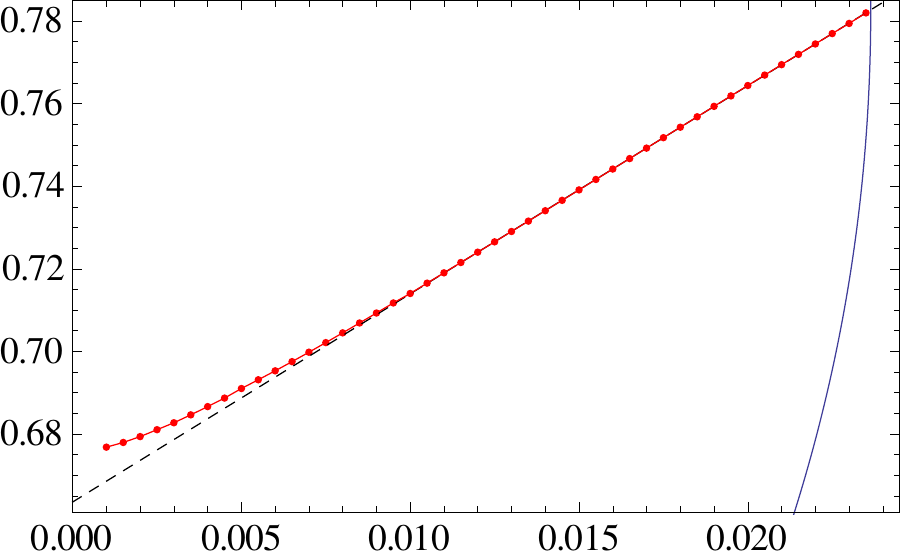}
\begin{picture}(0.1,0.1)(0,0)
\put(-115,-10){\makebox(0,0){$T/\mu$}}
\put(-245,85){\makebox(0,0){$k/\mu$}}
\end{picture}
\vskip 1em
\caption{The thermodynamically dominant striped black brane labelled by $k/\mu$ (\emph{i.e.} periodicity $2\pi\mu/k$) as a function of temperature $T/\mu$. The blue line is a section of the zero-mode curve displayed in figure \ref{bellcurve}. To accentuate the trend at low temperatures we have included the black dashed line which gives the best linear fit in the range $0.011<T/\mu<(T/\mu)_{c,\text{max}}$.\label{kt}}
\end{center}
\end{figure}
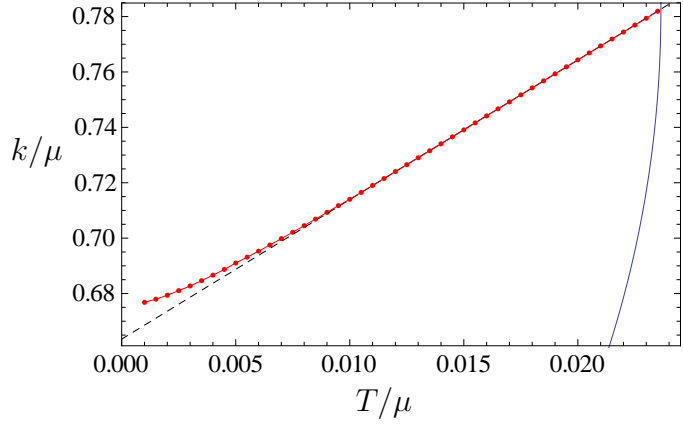

Extending this analysis to the two parameter family, in figure \ref{free3d} we show the averaged free energy $\bar{w}$ where the RN value has been subtracted. We have only found striped solutions below the threshold temperature, indicated in blue, where they dominate.\footnote{See however, an example of a different model in \cite{Withers:2013loa} where solutions at higher temperatures were found.}  Outside this region we have simply plotted the RN values.  Considering the system at any fixed $k/\mu$ we see that there is a continuous phase transition to the broken phase, as seen for a specific value in \cite{Withers:2013loa}. However, as emphasised we should determine the preferred $k/\mu$ as a function of the temperature. This locus of solutions is given by the red line in figure \ref{free3d}. For clarity we also plot the dominant $\tfrac{k}{\mu}(\frac{T}{\mu})$ in figure \ref{kt}, showing a monotonic growth in the stripe size as the temperature is reduced. Furthermore it approaches a non-zero value at low temperatures.

\begin{figure}[h!]
\begin{center}
\includegraphics[width=0.43\columnwidth]{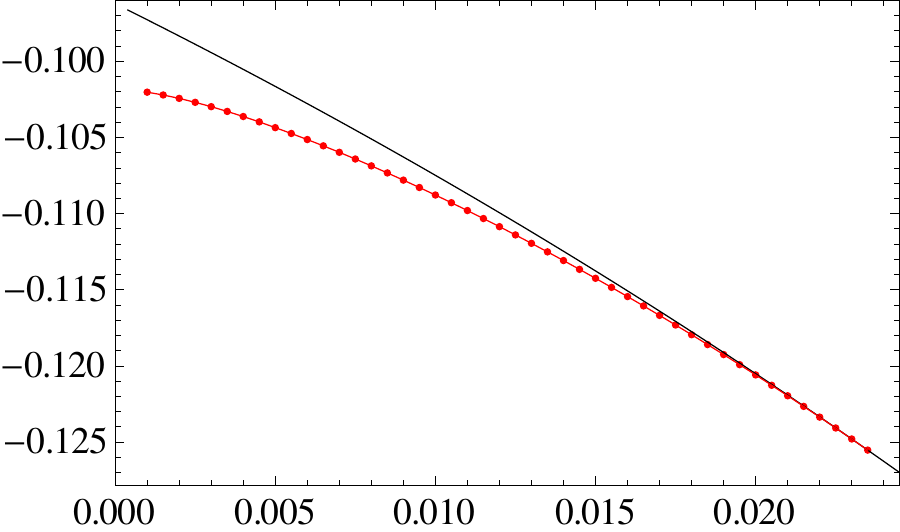}
\hspace{0.06\columnwidth}
\includegraphics[width=0.43\columnwidth]{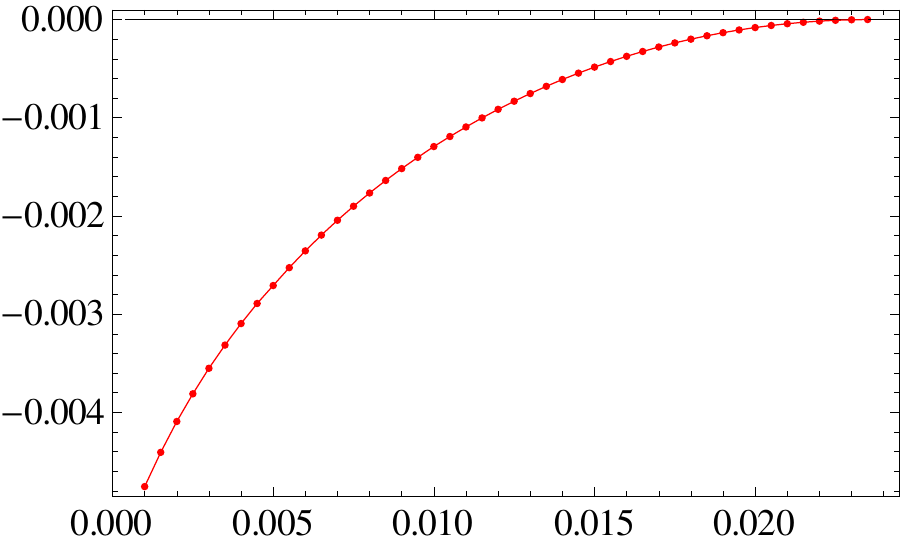}
\begin{picture}(0.1,0.1)(0,0)
\put(-440,80){\makebox(0,0){$\frac{\bar{w}}{\mu^3}$}}
\put(-215,80){\makebox(0,0){$\frac{\bar{w}-w_{\text{RN}}}{\mu^3}$}}
\put(-325,-10){\makebox(0,0){$T/\mu$}}
\put(-95,-10){\makebox(0,0){$T/\mu$}}
\end{picture}
\vskip 1em
\caption{\emph{Left:} Averaged free energy density for the thermodynamically preferred striped solutions labelled in figure \ref{kt}, with RN shown in black. \emph{Right:} Difference with the free energy density of the RN branch.\label{free2d}}
\end{center}
\end{figure}
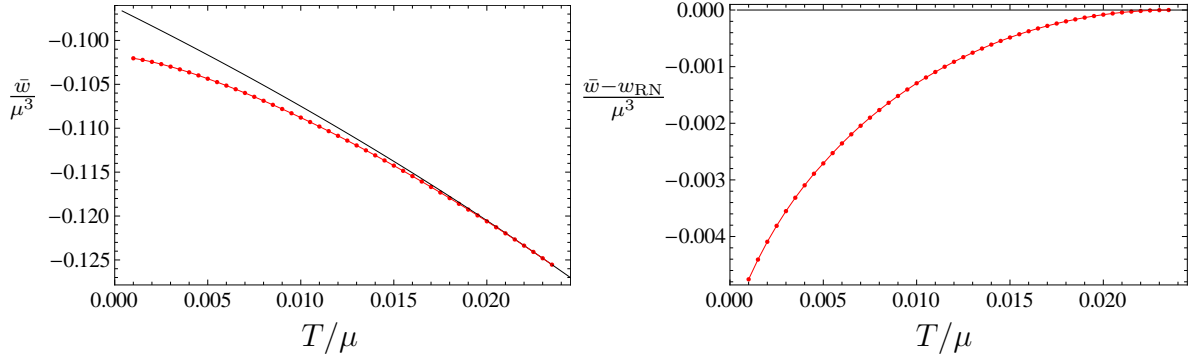

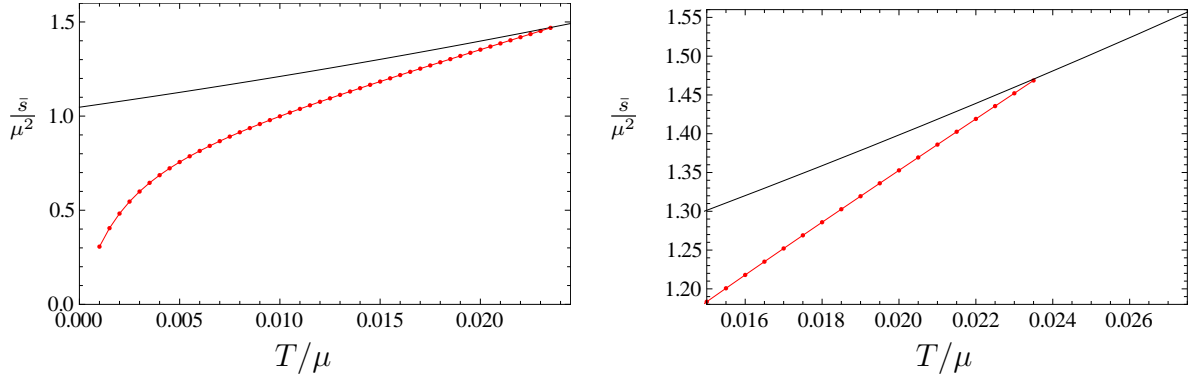
\begin{figure}[h!]
\begin{center}
\includegraphics[width=0.43\columnwidth]{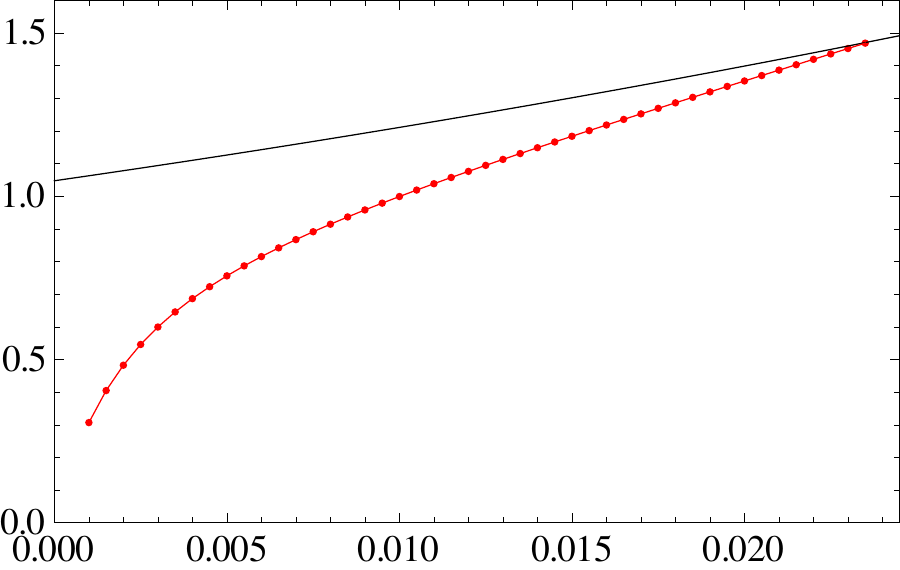}
\hspace{0.06\columnwidth}
\includegraphics[width=0.43\columnwidth]{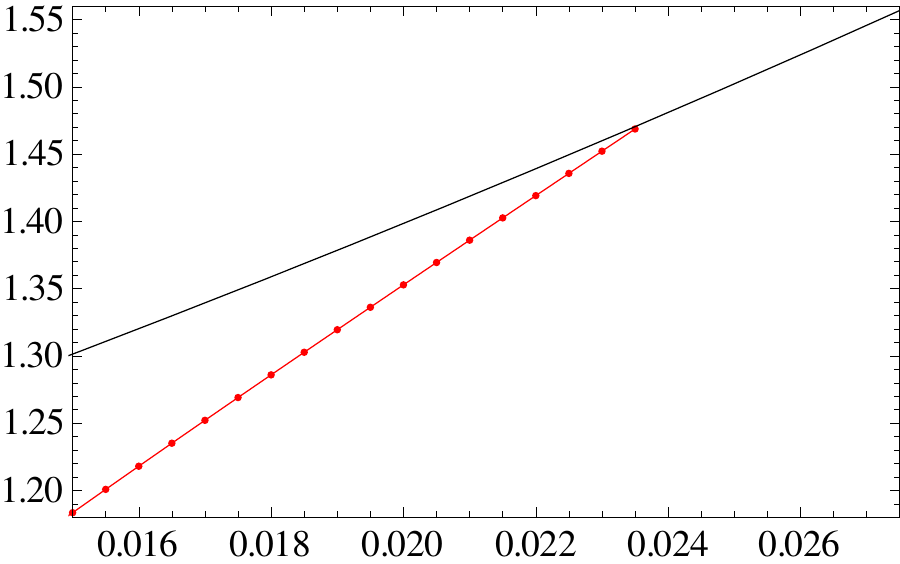}
\begin{picture}(0.1,0.1)(0,0)
\put(-440,80){\makebox(0,0){$\frac{\bar{s}}{\mu^2}$}}
\put(-215,80){\makebox(0,0){$\frac{\bar{s}}{\mu^2}$}}
\put(-335,-10){\makebox(0,0){$T/\mu$}}
\put(-95,-10){\makebox(0,0){$T/\mu$}}
\end{picture}
\vskip 1em
\caption{Averaged entropy density, $\bar{s}$, for the thermodynamically preferred striped solutions labelled in figure \ref{kt}. The right panel shows this quantity in the vicinity of the phase transition indicating a second order phase transition. RN is shown in black.\label{ent}}
\end{center}
\end{figure}

We now restrict our attention to this preferred line of solutions. Again, we see that there is a continuous phase transition, illustrated by the free energy shown in figure \ref{free2d}. Figure \ref{ent} shows the entropy density, which via the first law indicates a that the transition is second order. Following the system to lower temperatures, $T/\mu =  0.001$,  we see evidence of a zero entropy state emerging at zero temperature.

\begin{figure}[h!]
\begin{center}
\includegraphics[width=0.43\columnwidth]{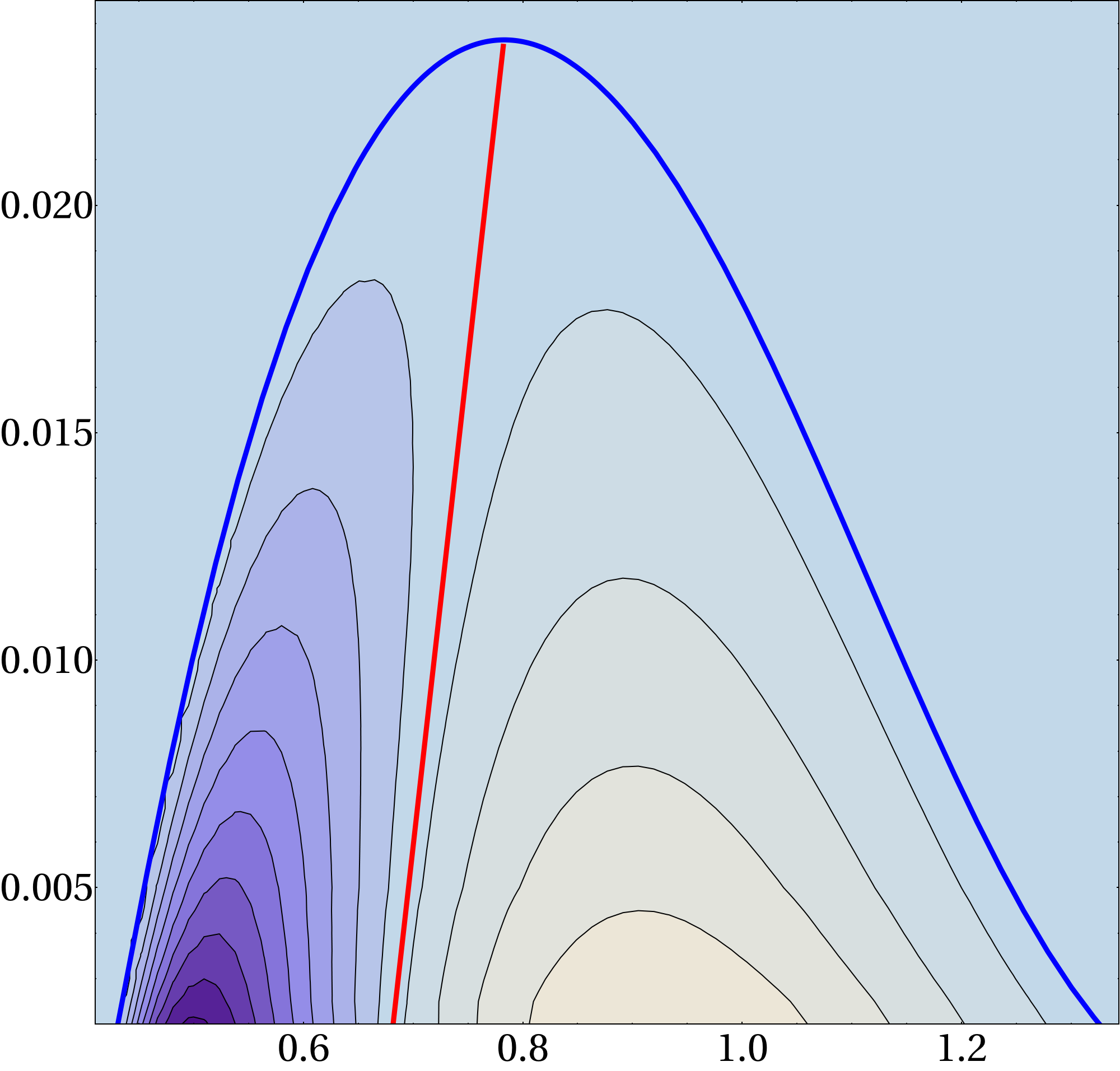}
\hspace{0.1\columnwidth}
\includegraphics[width=0.43\columnwidth]{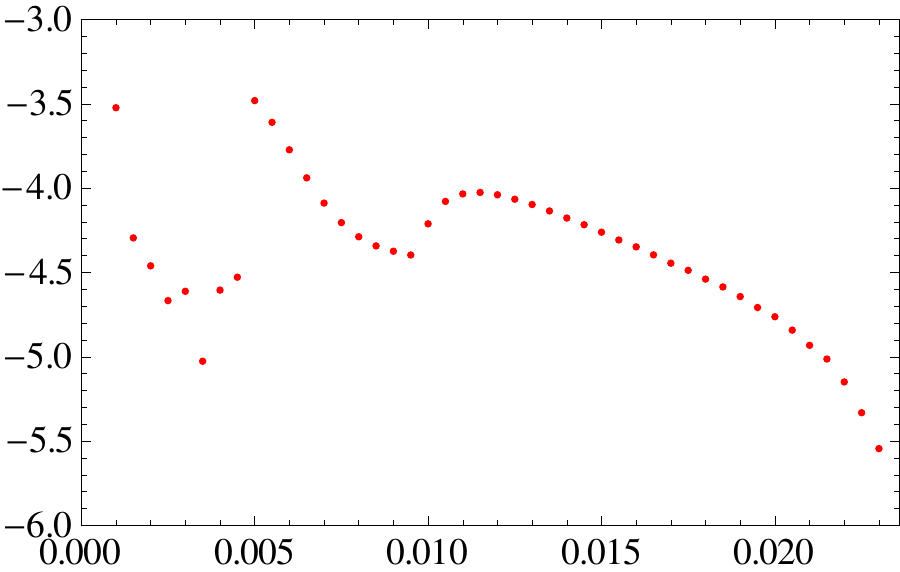}
\begin{picture}(0.1,0.1)(0,0)
\put(-350,-10){\makebox(0,0){$k/\mu$}}
\put(-450,133){\makebox(0,0){$T/\mu$}}
\put(-222,85){\makebox(0,0){$\log_{10}|\delta|$}}
\put(-100,-10){\makebox(0,0){$T/\mu$}}
\end{picture}
\vskip 1em
\caption{\emph{Left:} Contour plot for the quantity $\delta$ defined in \eqref{deltadef} for the two-parameter family of solutions. The blue curve shows the location of striped zero-modes about the RN solution. The red line shows the dominant striped solution at each $T/\mu$. \emph{Right:} The quantity $\log_{10}|\delta|$ for the thermodynamically preferred solutions showing $\delta \simeq 0$.  Note that the dominant error and associated scatter in this plot may be attributed to the extraction of the stress tensor; we observe convergence towards zero with $N$. \label{delta3d}}
\end{center}
\end{figure}
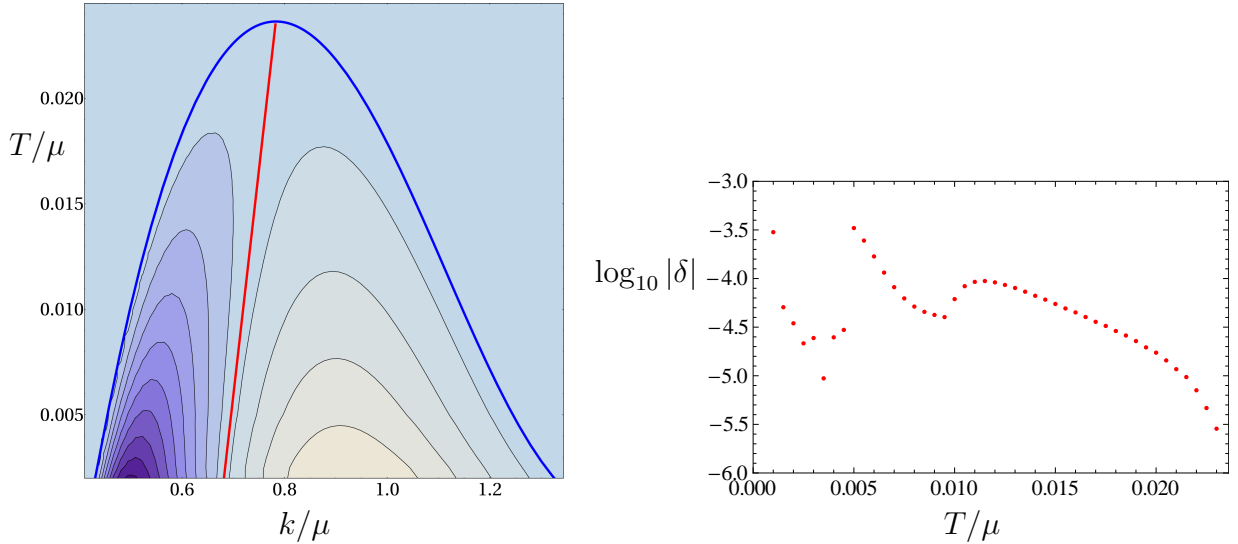

For the five-dimensional helical black branes constructed in \cite{Donos:2012gg,Donos:2012wi}, it was found that the thermodynamically preferred solutions at fixed temperature were characterised by the vanishing of a particular quantity in the boundary stress tensor. Here we work in four dimensions, and we seek a analogous statement. One way of characterising the vanishing quantity of \cite{Donos:2012gg,Donos:2012wi} is through the relation $\left<T_{xx}\right> = \bar{\varepsilon}/d$ in the case where the stress tensor depends on the coordinate $x$ and $d$ is the spatial dimension of the boundary.\footnote{In the notation of  \cite{Donos:2012gg,Donos:2012wi}, $x_\text{here} = x_1$ whilst the relation $\left<T_{xx}\right> = \bar{\varepsilon}/d$ amounts to the vanishing of the boundary data $c_h$.}
Indeed, we find that along the preferred line of $d=2$ inhomogeneous solutions studied in this paper,  this relation holds within numerical accuracy, and is not satisfied away from this locus. This is clearly demonstrated for the space of solutions at the fixed $T/\mu$ shown in figure \ref{fixedT}. This is equivalent to the relation amongst averaged boundary data, $2X^{(3)}+\bar{T}^{(3)}=0$. For all solutions presented in this paper we find that the spatial average of $\left<T_{ty}\right>$ and $\left<J_y\right>$ vanishes, consistent with the perturbative structure \cite{Donos:2011bh}. Consequently if the relation $\left<T_{xx}\right>= \bar{\varepsilon}/d$ is satisfied, then conformality implies that both the spatially averaged stress tensor and spatially averaged current are isotropic. In order to numerically test this relation more comprehensively, it is convenient to define the dimensionless ratio,
\be
\delta \equiv \frac{\left<T_{xx}\right> - \bar{\varepsilon}/d}{\left<T_{xx}\right>}.\label{deltadef}
\ee
In figure \ref{delta3d} we plot this quantity for the two parameter family of black branes. In general, $\delta\neq 0$, but we find that it vanishes along the line of thermodynamically preferred solutions.

\section{Final comments}
We have constructed a two-parameter family of cohomogeneity-two black brane solutions with AdS$_4$ asymptotics, dual to a phase at finite charge density with spontaneously broken translational invariance in one direction. The family may be described as a space of solutions labelled by their periodicity in one of the spatial boundary directions, $2\pi\mu/k$, at each temperature $T/\mu$. This work builds on the results \cite{Withers:2013loa} (see also \cite{DonosSolo}) where second order phase transitions were found for one parameter families at fixed $k/\mu$ in this model.

In section \ref{nonlinear}, for the two parameter family, we find continuous phase transitions at the threshold of the linear striped instabilities. In other words, if we were to fix the periodicity of the system somehow, then the phase boundary would be given by the zero-mode analysis, illustrated by the characteristic `bell curve' in figure \ref{bellcurve}. However, in the absence of such a restriction we are interested in those solutions  which minimise the free energy at a given temperature, $T/\mu$. Constructing this set of solutions, we have found the preferred scale as a function of temperature $\tfrac{k}{\mu}(\tfrac{T}{\mu})$, with the stripe size monotonically growing as the temperature is reduced, approaching a non-zero value at low temperatures.

We also examined the properties of this preferred set of solutions. As in the fixed $k/\mu$ case, we have shown that the system exhibits a second order phase transition to the inhomogeneous phase, with the entropy appearing to vanish as the inhomogeneous zero temperature state is approached. Constructing these states directly at zero temperature would be profitable. Motivated by an analogous feature in the helical black brane setting \cite{Donos:2012gg,Donos:2012wi} we have examined a particular combination of stress-tensor components, $\delta$. We find that whilst $\delta$ is non-zero in general for the two parameter family, it vanishes along the line of solutions which dominate the ensemble at fixed $T/\mu$. In this case, vanishing $\delta$ implies isotropy of the spatially-averaged stress tensor. It would be interesting to investigate the origin of this feature.

Finally, we stress that we have focussed on cohomogeneity-two solutions. More generally we anticipate that there is a cohomogeneity-three family of solutions emerging from the bell-curve (figure \ref{bellcurve}), which may turn out to dominate the ensemble.

\section*{Acknowledgements}
We thank Aristomenis Donos, Jerome Gauntlett, Julian Sonner and Toby Wiseman for useful discussions. We thank the Perimeter Institute and Nordita for hospitality.
This work is supported by a Royal Commission for the Exhibition of 1851 Science Research Fellowship.
\bibliographystyle{utphys}
\bibliography{stripes}{}

\providecommand{\href}[2]{#2}\begingroup\raggedright\begin{thebibliography}{10}

\bibitem{Gubser:2008px}
S.~S. Gubser, ``{Breaking an Abelian Gauge Symmetry Near a Black Hole
  Horizon},'' \href{http://dx.doi.org/10.1103/PhysRevD.78.065034}{{\em Phys.
  Rev.} {\bfseries D78} (2008) 065034},
\href{http://arxiv.org/abs/0801.2977}{{\ttfamily arXiv:0801.2977 [hep-th]}}.

\bibitem{Hartnoll:2008vx}
S.~A. Hartnoll, C.~P. Herzog, and G.~T. Horowitz, ``{Building a Holographic
  Superconductor},''
  \href{http://dx.doi.org/10.1103/PhysRevLett.101.031601}{{\em Phys. Rev.
  Lett.} {\bfseries 101} (2008) 031601},
\href{http://arxiv.org/abs/0803.3295}{{\ttfamily arXiv:0803.3295 [hep-th]}}.

\bibitem{Hartnoll:2008kx}
S.~A. Hartnoll, C.~P. Herzog, and G.~T. Horowitz, ``{Holographic
  Superconductors},''
  \href{http://dx.doi.org/10.1088/1126-6708/2008/12/015}{{\em JHEP} {\bfseries
  12} (2008) 015},
\href{http://arxiv.org/abs/0810.1563}{{\ttfamily arXiv:0810.1563 [hep-th]}}.

\bibitem{Gauntlett:2009dn}
J.~P. Gauntlett, J.~Sonner, and T.~Wiseman, ``{Holographic superconductivity in
  M-Theory},'' \href{http://dx.doi.org/10.1103/PhysRevLett.103.151601}{{\em
  Phys. Rev. Lett.} {\bfseries 103} (2009) 151601},
\href{http://arxiv.org/abs/0907.3796}{{\ttfamily arXiv:0907.3796 [hep-th]}}.

\bibitem{Gubser:2009qm}
S.~S. Gubser, C.~P. Herzog, S.~S. Pufu, and T.~Tesileanu, ``{Superconductors
  from Superstrings},''
  \href{http://dx.doi.org/10.1103/PhysRevLett.103.141601}{{\em Phys. Rev.
  Lett.} {\bfseries 103} (2009) 141601},
\href{http://arxiv.org/abs/0907.3510}{{\ttfamily arXiv:0907.3510 [hep-th]}}.

\bibitem{Nakamura:2009tf}
S.~Nakamura, H.~Ooguri, and C.-S. Park, ``{Gravity Dual of Spatially Modulated
  Phase},'' \href{http://dx.doi.org/10.1103/PhysRevD.81.044018}{{\em Phys.
  Rev.} {\bfseries D81} (2010) 044018},
\href{http://arxiv.org/abs/0911.0679}{{\ttfamily arXiv:0911.0679 [hep-th]}}.

\bibitem{Ooguri:2010kt}
H.~Ooguri and C.-S. Park, ``{Holographic End-Point of Spatially Modulated Phase
  Transition},'' \href{http://dx.doi.org/10.1103/PhysRevD.82.126001}{{\em Phys.
  Rev.} {\bfseries D82} (2010) 126001},
\href{http://arxiv.org/abs/1007.3737}{{\ttfamily arXiv:1007.3737 [hep-th]}}.

\bibitem{Donos:2011bh}
A.~Donos and J.~P. Gauntlett, ``{Holographic striped phases},''
  \href{http://dx.doi.org/10.1007/JHEP08(2011)140}{{\em JHEP} {\bfseries 1108}
  (2011) 140},
\href{http://arxiv.org/abs/1106.2004}{{\ttfamily arXiv:1106.2004 [hep-th]}}.

\bibitem{Donos:2011ff}
A.~Donos and J.~P. Gauntlett, ``{Holographic helical superconductors},''
  \href{http://dx.doi.org/10.1007/JHEP12(2011)091}{{\em JHEP} {\bfseries 12}
  (2011) 091},
\href{http://arxiv.org/abs/1109.3866}{{\ttfamily arXiv:1109.3866 [hep-th]}}.

\bibitem{Vegh:2013sk}
D.~Vegh, ``{Holography without translational symmetry},''
\href{http://arxiv.org/abs/1301.0537}{{\ttfamily arXiv:1301.0537 [hep-th]}}.

\bibitem{Donos:2013gda}
A.~Donos and J.~P. Gauntlett, ``{Holographic charge density waves},''
\href{http://arxiv.org/abs/1303.4398}{{\ttfamily arXiv:1303.4398 [hep-th]}}.

\bibitem{Cremonini:2012ir}
S.~Cremonini and A.~Sinkovics, ``{Spatially Modulated Instabilities of
  Geometries with Hyperscaling Violation},''
\href{http://arxiv.org/abs/1212.4172}{{\ttfamily arXiv:1212.4172 [hep-th]}}.

\bibitem{Iizuka:2013ag}
N.~Iizuka and K.~Maeda, ``{Stripe Instabilities of Geometries with Hyperscaling
  Violation},''
\href{http://arxiv.org/abs/1301.5677}{{\ttfamily arXiv:1301.5677 [hep-th]}}.

\bibitem{Domokos:2007kt}
S.~K. Domokos and J.~A. Harvey, ``{Baryon number-induced Chern-Simons couplings
  of vector and axial-vector mesons in holographic QCD},''
  \href{http://dx.doi.org/10.1103/PhysRevLett.99.141602}{{\em Phys. Rev. Lett.}
  {\bfseries 99} (2007) 141602},
\href{http://arxiv.org/abs/0704.1604}{{\ttfamily arXiv:0704.1604 [hep-ph]}}.

\bibitem{Ooguri:2010xs}
H.~Ooguri and C.-S. Park, ``{Spatially Modulated Phase in Holographic
  Quark-Gluon Plasma},''
  \href{http://dx.doi.org/10.1103/PhysRevLett.106.061601}{{\em Phys. Rev.
  Lett.} {\bfseries 106} (2011) 061601},
\href{http://arxiv.org/abs/1011.4144}{{\ttfamily arXiv:1011.4144 [hep-th]}}.

\bibitem{Bayona:2011ab}
C.~A.~B. Bayona, K.~Peeters, and M.~Zamaklar, ``{A non-homogeneous ground state
  of the low-temperature Sakai-Sugimoto model},''
  \href{http://dx.doi.org/10.1007/JHEP06(2011)092}{{\em JHEP} {\bfseries 06}
  (2011) 092},
\href{http://arxiv.org/abs/1104.2291}{{\ttfamily arXiv:1104.2291 [hep-th]}}.

\bibitem{Bergman:2011rf}
O.~Bergman, N.~Jokela, G.~Lifschytz, and M.~Lippert, ``{Striped instability of
  a holographic Fermi-like liquid},''
  \href{http://dx.doi.org/10.1007/JHEP10(2011)034}{{\em JHEP} {\bfseries 10}
  (2011) 034},
\href{http://arxiv.org/abs/1106.3883}{{\ttfamily arXiv:1106.3883 [hep-th]}}.

\bibitem{Iizuka:2012iv}
N.~Iizuka, S.~Kachru, N.~Kundu, P.~Narayan, N.~Sircar, {\em et al.}, ``{Bianchi
  Attractors: A Classification of Extremal Black Brane Geometries},''
\href{http://arxiv.org/abs/1201.4861}{{\ttfamily arXiv:1201.4861 [hep-th]}}.

\bibitem{Donos:2012gg}
A.~Donos and J.~P. Gauntlett, ``{Helical superconducting black holes},''
  \href{http://dx.doi.org/10.1103/PhysRevLett.108.211601}{{\em Phys.Rev.Lett.}
  {\bfseries 108} (2012) 211601},
\href{http://arxiv.org/abs/1203.0533}{{\ttfamily arXiv:1203.0533 [hep-th]}}.

\bibitem{Donos:2012wi}
A.~Donos and J.~P. Gauntlett, ``{Black holes dual to helical current phases},''
\href{http://arxiv.org/abs/1204.1734}{{\ttfamily arXiv:1204.1734 [hep-th]}}.

\bibitem{DonosSolo}
A.~Donos, ``{Striped phases from holography},''
\href{http://arxiv.org/abs/1303.7211}{{\ttfamily arXiv:1303.7211 [hep-th]}}.

\bibitem{Withers:2013loa}
B.~Withers, ``{Black branes dual to striped phases},''
\href{http://arxiv.org/abs/1304.0129}{{\ttfamily arXiv:1304.0129 [hep-th]}}.

\bibitem{UBCv1}
M.~Rozali, D.~Smyth, E.~Sorkin, and J.~B. Stang, ``{Holographic Stripes},''
\href{http://arxiv.org/abs/1211.5600v1}{{\ttfamily arXiv:1211.5600v1
  [hep-th]}}.

\bibitem{UBCv2}
M.~Rozali, D.~Smyth, E.~Sorkin, and J.~B. Stang, ``{Holographic Stripes},''
\href{http://arxiv.org/abs/1211.5600v2}{{\ttfamily arXiv:1211.5600v2
  [hep-th]}}.

\bibitem{Gauntlett:2009bh}
J.~P. Gauntlett, J.~Sonner, and T.~Wiseman, ``{Quantum Criticality and
  Holographic Superconductors in M- theory},''
  \href{http://dx.doi.org/10.1007/JHEP02(2010)060}{{\em JHEP} {\bfseries 02}
  (2010) 060},
\href{http://arxiv.org/abs/0912.0512}{{\ttfamily arXiv:0912.0512 [hep-th]}}.

\bibitem{Gauntlett:2009zw}
J.~P. Gauntlett, S.~Kim, O.~Varela, and D.~Waldram, ``{Consistent
  Supersymmetric Kaluza--Klein Truncations with Massive Modes},''
  \href{http://dx.doi.org/10.1088/1126-6708/2009/04/102}{{\em JHEP} {\bfseries
  04} (2009) 102},
\href{http://arxiv.org/abs/0901.0676}{{\ttfamily arXiv:0901.0676 [hep-th]}}.

\bibitem{Donos:2012yu}
A.~Donos, J.~P. Gauntlett, J.~Sonner, and B.~Withers, ``{Competing orders in
  M-theory: superfluids, stripes and metamagnetism},''
  \href{http://dx.doi.org/10.1007/JHEP03(2013)108}{{\em JHEP} {\bfseries 1303}
  (2013) 108},
\href{http://arxiv.org/abs/1212.0871}{{\ttfamily arXiv:1212.0871 [hep-th]}}.

\bibitem{Balasubramanian:1999re}
V.~Balasubramanian and P.~Kraus, ``{A stress tensor for anti-de Sitter
  gravity},'' \href{http://dx.doi.org/10.1007/s002200050764}{{\em Commun. Math.
  Phys.} {\bfseries 208} (1999) 413--428},
\href{http://arxiv.org/abs/hep-th/9902121}{{\ttfamily arXiv:hep-th/9902121}}.

\bibitem{Headrick:2009pv}
M.~Headrick, S.~Kitchen, and T.~Wiseman, ``{A New approach to static numerical
  relativity, and its application to Kaluza-Klein black holes},''
  \href{http://dx.doi.org/10.1088/0264-9381/27/3/035002}{{\em
  Class.Quant.Grav.} {\bfseries 27} (2010) 035002},
\href{http://arxiv.org/abs/0905.1822}{{\ttfamily arXiv:0905.1822 [gr-qc]}}.

\bibitem{Adam:2011dn}
A.~Adam, S.~Kitchen, and T.~Wiseman, ``{A numerical approach to finding general
  stationary vacuum black holes},''
  \href{http://dx.doi.org/10.1088/0264-9381/29/16/165002}{{\em
  Class.Quant.Grav.} {\bfseries 29} (2012) 165002},
\href{http://arxiv.org/abs/1105.6347}{{\ttfamily arXiv:1105.6347 [gr-qc]}}.

\bibitem{Wiseman:2011by}
T.~Wiseman, ``{Numerical construction of static and stationary black holes},''
\href{http://arxiv.org/abs/1107.5513}{{\ttfamily arXiv:1107.5513 [gr-qc]}}.

\bibitem{Horowitz:2012ky}
G.~T. Horowitz, J.~E. Santos, and D.~Tong, ``{Optical Conductivity with
  Holographic Lattices},''
  \href{http://dx.doi.org/10.1007/JHEP07(2012)168}{{\em JHEP} {\bfseries 1207}
  (2012) 168},
\href{http://arxiv.org/abs/1204.0519}{{\ttfamily arXiv:1204.0519 [hep-th]}}.

\bibitem{Horowitz:2012gs}
G.~T. Horowitz, J.~E. Santos, and D.~Tong, ``{Further Evidence for
  Lattice-Induced Scaling},''
  \href{http://dx.doi.org/10.1007/JHEP11(2012)102}{{\em JHEP} {\bfseries 1211}
  (2012) 102},
\href{http://arxiv.org/abs/1209.1098}{{\ttfamily arXiv:1209.1098 [hep-th]}}.

\bibitem{Horowitz:2013jaa}
G.~T. Horowitz and J.~E. Santos, ``{General Relativity and the Cuprates},''
\href{http://arxiv.org/abs/1302.6586}{{\ttfamily arXiv:1302.6586 [hep-th]}}.

\end{thebibliography}\endgroup
\end{document}